# Modelling & Steady State Compliance Testing of an Improved Time Synchronized Phasor Measurement Unit Based on IEEE Standard C37.118.1

Nisarg Trivedi
*Department of Electrical Engineering*
*S. V. National Institute of Technology – SVNIT*
Surat, India
nisarg819@gmail.com

*Abstract*— Synchrophasor technology is an emerging and developing technology for monitoring and control of wide area measurement systems (WAMS). In an elementary WAMS, two identical phasors measured at two different locations have difference in the phase angles measured since their reference waveforms are not synchronized with each other. Phasor measurement units (PMUs) measure input phasors with respect to a common reference wave based on the atomic clock pulses received from global positioning system (GPS) satellites, eliminating variation in the measured phase angles due to distant locations of the measurement nodes. This has found tremendous applications in quick fault detection, fault location analysis, accurate current, voltage, frequency and phase angle measurements in WAMS. Commercially available PMU models are often proven to be expensive for research and development as well as for grid integration projects. This research article proposes an economic PMU model optimized for accurate steady-state performance based on recursive discrete Fourier transform (DFT) and provides results and detailed analysis of the proposed PMU model as per the steady state compliance specifications of IEEE standard C37.118.1. Results accurate up to 13 digits after decimal point are obtained through the developed PMU model for both nominal and off-nominal frequency inputs in steady state.

## I. INTRODUCTION

THIS research article provides a detailed modelling and steady state compliance testing of a time synchronized phasor measurement unit based on recursive discrete Fourier transform (DFT) technique. Several approaches to model a PMU have been adopted and implemented in detail over the last two decades, such as zero-crossing detection method[1], Kalman filtering method [2], discrete Fourier transform (DFT) techniques[3], quadrature demodulation technique[4], Z-transform and median filtering technique[5], Newton-type approximations[6], phase-locked-loop (PLL) technique[7] and Prony's method[8]. Existing literature also proposes improvements and modifications to the synchrophasor estimation using DFT techniques [9]. Recursive DFT technique is mostly preferred for implementing a low-cost PMU considering its simple, straight-forward algorithm and economic implementation. Existing literature in this field discusses several PMU models based on recursive DFT technique majorly focused on its analysis over dynamic conditions, such as [10, 11]. However, none of them provides a detailed PMU model based on recursive DFT technique considering utmost accuracy in steady state conditions which is provided in this research article. PMUs, chiefly used for the time synchronized phasor measurement provide several advantages over conventional phasor measurement devices such as accurate and fast fault identification and protection initialisation, accurate fault location determination, correct phasor measurement irrespective of location of measurement system etc. Due to these advantages, PMUs have found tremendous applications in smart grids, micro grids, distribution grid management & control systems and in WAMS [12]. Existing research literature is majorly focused over performance of PMUs under dynamic conditions. However, with modern technological improvements in operation and control of power systems, the system parameters are more rigidly and ruggedly controlled, or in other words, the power system is becoming more robust against disturbances with advancement in control technology.

Under such circumstances, it becomes equally important to study performance of PMU under steady state conditions and develop new PMU models to achieve more accurate results for better reliability of the measurement system. Recursive DFT technique is mainly used for implementing a PMU as it is an economic and relatively less complex algorithm. This technique has been adopted for the proposed PMU model with certain modifications improving accuracy of the PMU phasor estimates for both nominal and off-nominal frequency inputs while also maintaining speed of PMU response. The performance of the developed model has been verified against compliance requirements for steady state specified in IEEE standard C37.118.1 [13, 14].

## II. RECURSIVE DFT TECHNIQUE FOR NOMINAL AND OFF-NOMINAL FREQUENCY INPUTS [15]

The proposed PMU model has been developed based on recursive DFT technique with certain modifications for greater accuracy in steady state performance. Equations related with recursive DFT technique for nominal and off-nominal frequency inputs are taken from the reference [15]. The input signal is considered to be

$$x(t) = X_m \cos(2\pi f_0 t + \phi) \qquad (1)$$

Where,

$f_0$ = nominal frequency of power system (50 Hz)

$\phi$ = phase angle of the input signal

$X_m$ = maximum amplitude of the input signal

If $x(t)$ is sampled N times over each periodic cycle of angular period $2\pi$, then sampled values of $x(t)$ are represented as

$$x(n) = X_m \cos(\frac{2\pi}{N}n + \phi) \qquad (2)$$

For simplicity, let $\theta = \frac{2\pi}{N}$. Therefore,

$$x(n) = X_m \cos(n\theta + \phi) \qquad (3)$$

On applying Discrete Fourier Transform (DFT), we get magnitudes of various frequency components present in the signal x(n) as:

$$X[k] = \frac{\sqrt{2}}{N}\sum_{n=0}^{N-1} x(n)e^{-j\theta nk} \qquad (4)$$

Where k is index of frequency component. In all following calculations, k=1 is considered for fundamental frequency calculations. Simplified form of (4) is used to calculate samples of the first window with n=0 to n=N-1. The calculation for consecutive windows is simply repeated through following equations. This is termed as non-recursive DFT technique.

$$X^{N-1} = \frac{\sqrt{2}}{N}\sum_{n=0}^{N-1} x(n)[\cos(n\theta) - j\sin(n\theta)]$$

$$X^{N} = \frac{\sqrt{2}}{N}\sum_{n=0}^{N-1} x(n+1)[\cos(n\theta) - j\sin(n\theta)] \qquad (5)$$

These equations can be expressed in phasor form as

$$X^{N-1} = \frac{\sqrt{2}}{N}\sum_{n=0}^{N-1} x(n)e^{-jn\theta}$$

$$X^{N} = \frac{\sqrt{2}}{N}\sum_{n=0}^{N-1} x(n+1)e^{-jn\theta} \qquad (6)$$

First window runs for n=0 to n=N-1; while second window runs for n=1 to n=N. The factor $e^{-j\theta}$ is multiplied at both sides of (5) to achieve the following result:

$$\hat{X}^N = e^{-j\theta} X^N = \frac{\sqrt{2}}{N}\sum_{n=0}^{N-1} x(n+1)e^{-j(n+1)\theta}$$

$$= X^{N-1} + \frac{\sqrt{2}}{N}(x(N) - x(0))e^{-j(0)\theta} \qquad (7)$$

Here the identity $e^{-j(0)\theta} = 1 = e^{-jN\theta}$ has been used since one period over fundamental frequency is constituted by exactly N samples. Since (N-1) samples of new window are identical with that of the previous window, considerable time and computations are reduced. This is known as the recursive DFT algorithm [15]. In general, for a data window with last sample as $(N+r)^{th}$, the phasor estimate can be calculated recursively as

$$\hat{X}^{N+r} = e^{-j\theta} X^{N+r-1} = \frac{\sqrt{2}}{N}(x(n+r) - x(r))e^{-jr\theta}$$

$$= \hat{X}^{N+r-1} + \frac{\sqrt{2}}{N}(x(N+r) - x(r))e^{-j(r)\theta} \qquad (8)$$

From (8), for a power system with nominal frequency sinusoid, $x(N+r) = x(r)$. Now, let's consider a case of off-nominal angular frequency $\omega$ given as

$$\omega = \omega_0 + \Delta\omega \qquad (9)$$

The input signal is:

$$x(t) = X_m \cos(\omega t + \phi)$$

$$= \sqrt{2}\,\text{Re}[\frac{X_m}{\sqrt{2}} e^{j\phi} e^{j\omega t}] \qquad (10)$$

$$= \sqrt{2}\,\text{Re}[Xe^{j\omega t}]$$

Where $X$ = correct phasor estimate at off-nominal frequency and Re suggests real value function. Equation (10) can be expressed as

$$x(t) = \frac{\sqrt{2}}{2}[Xe^{j\omega t} + X^* e^{-j\omega t}] \qquad (11)$$

While converting this signal in discrete form, $k^{th}$ sample is given by

$$x(k) = \frac{1}{\sqrt{2}}[Xe^{j\omega k\Delta t} + X^* e^{-j\omega k\Delta t}] \qquad (12)$$

For an off-nominal frequency signal $x(t)$, its phasor estimate is denoted by $X'$ which is calculated using (8) keeping the first sample x(r). Thus $X'_r$ can be given by

$$X'_r = \frac{\sqrt{2}}{N}\sum_{k=r}^{r+N-1} x(k)e^{-j\omega_0 k\Delta t}$$

$$= \frac{1}{N}\sum_{k=r}^{r+N-1}[Xe^{j\omega k\Delta t} + X^* e^{-j\omega k\Delta t}]e^{-j\omega_0 k\Delta t} \qquad (13)$$

Since,

$$e^{jx} - 1 = e^{jx/2}(e^{jx/2} - e^{-jx/2})$$

$$= 2je^{jx/2}\sin(x/2) \qquad (14)$$

The two terms added together in (13) are basically geometric series which can be rewritten in closed form as shown in (14). The mathematical steps involved in this simplification are given in [16].

$$X'_r = Xe^{jr(\omega-\omega_0)\Delta t}\left\{\frac{\sin(\frac{N(\omega-\omega_0)\Delta t}{2})}{N\sin(\frac{(\omega-\omega_0)\Delta t}{2})}\right\}e^{+j(N-1)\frac{(\omega-\omega_0)\Delta t}{2}} +$$

$$X^* e^{-jr(\omega+\omega_0)\Delta t}\left\{\frac{\sin(\frac{N(\omega+\omega_0)\Delta t}{2})}{N\sin(\frac{(\omega+\omega_0)\Delta t}{2})}\right\}e^{-j(N-1)\frac{(\omega+\omega_0)\Delta t}{2}}$$

$$(15)$$

Or, $X'_r = PXe^{jr(\omega-\omega_0)\Delta t} + QX^* e^{-jr(\omega+\omega_0)\Delta t} \qquad (16)$

Where P and Q are the coefficients independent of 'r', which are given by (17):

$$P = \left\{\frac{\sin(\frac{N(\omega-\omega_0)\Delta t}{2})}{N\sin(\frac{(\omega-\omega_0)\Delta t}{2})}\right\}e^{+j(N-1)\frac{(\omega-\omega_0)\Delta t}{2}},$$

$$Q = \left\{\frac{\sin(\frac{N(\omega+\omega_0)\Delta t}{2})}{N\sin(\frac{(\omega+\omega_0)\Delta t}{2})}\right\}e^{-j(N-1)\frac{(\omega+\omega_0)\Delta t}{2}} \qquad (17)$$

The phase angle for an off-nominal frequency phasor estimate rotates from its initial value to a complete cycle of 2π radians at a period determined by the difference between off-nominal and nominal frequency.

Period of phasor rotation = $T = \frac{1}{|f_{in} - f_0|} \qquad (18)$

Where, $f_{in}$ = off-nominal input frequency
$f_0$ = nominal frequency (50 Hz)

III. IMPLEMENTATION OF THE PMU MODEL

The PMU model is developed as a MATLAB script based on above equations of the recursive DFT technique. The function takes as input the absolute time pulse 't' and an unknown input signal 'v'. The function calculates RMS amplitude, frequency, phase angle, sample frequency, sampling period N and rate of change of frequency (ROCOF) for the unknown input signal 'v' given.
The detailed algorithm is as following:
1. RMS amplitude of the input signal is found at first. Number of samples within a complete period of a signal is found using zero

crossing detection. Within each such period, the RMS magnitude of the signal is found using the following equations:

$$V_{RMS}(m) = \sqrt{\frac{\sum_{n=m}^{n=m+T} v(n)^2}{T}} \quad (19)$$

Where,

$V_{RMS}(m)$ = instantaneous RMS amplitude of input signal v at $m^{th}$ sample

$T$ = discrete time period of signal v detected by zero crossing detection technique

To increase accuracy of RMS amplitude, a 15 point moving average of instantaneous samples is carried out as

$$V_{RMS}(m) = \frac{\sum_{n=m-14}^{n=m} V_{RMS}(n)}{15} \quad (20)$$

2. To find frequency of the unknown signal v, first it is divided by maximum amplitude $V_{max}$ to get cosine envelope. The frequency is then found by detecting change in angles for two consecutive time samples.

$$u(n) = \frac{v(n)}{V_{max}(n)} = \cos(2\pi t(n) + \varphi) \quad (21)$$

Where, $V_{max}(n) = V_{RMS}(n) * \sqrt{2}$ (22)

$$f(n) = \frac{\cos^{-1} u(n) - \cos^{-1} u(n-1)}{2\pi(t(n) - t(n-1))} \quad (23)$$

Similar to amplitude calculation, moving average of 20 samples is carried out to achieve better accuracy.

3. Input sampling frequency $F_s$ is found by detecting difference between two consecutive time pulses.

$$F_s(n) = \frac{1}{t(n) - t(n-1)} \quad (24)$$

Discrete time sampling period N is found by representing the ratio of $F_0$ and $F_s$ as ratio of two integers and considering the denominator to be N. For example, if $F_s$=10000 and $F_0$=50, then N = denominator of $\left(\frac{F_0}{F_s}\right)$= 200.

4. Rate of change of frequency (ROCOF) is found using the discrete form equation

$$ROCOF(n) = \frac{F(n) - F(n-1)}{t(n) - t(n-1)} \quad (25)$$

5. For a nominal frequency input, the phasor estimate of the first cycle is calculated using equation (6) and for consecutive windows, using the equation (8).

6. For an off- nominal frequency input, the phasor estimate of the first window is calculated using equations (16) & (17) and then corrected against the equation (18) to get the most accurate results, assuming purely cosine wave of a constant off-nominal frequency for one window. For the consecutive windows, the new phasor estimates are calculated recursively using conventional equations of (16) & (17).

This algorithm calculates phasor estimates for both nominal and off-nominal frequency inputs and provides output accurate up to 13 digits after decimal point for input voltage v up to as high as 11 kV.

IV. PMU RESULTS FOR NOMINAL AND OFF-NOMINAL FREQUENCY INPUTS

A. *PMU results for nominal frequency inputs:*

The developed PMU model has been tested for 4 nominal frequency (50 Hz) inputs with different phase angles. These testing inputs are unknown to the PMU when they are fed to it. The PMU very accurately and precisely determines the magnitude, phase angle and frequency of each unknown input phasor. The input phasor can be either voltage or current – PMU treats both of them identically in digital form. The 4 testing inputs with nominal frequency (50 Hz) and different phase angles are:

Input 1: $V = 230 * \sqrt{2} \cos(\omega_0 t - \frac{\pi}{6})$

Input 2: $V = 230 * \sqrt{2} \cos(\omega_0 t + 0)$

Input 3: $V = 230 * \sqrt{2} \cos(\omega_0 t + \frac{\pi}{6})$

Input 4: $V = 230 * \sqrt{2} \cos(\omega_0 t - \pi)$

Where, $\omega_0 = 2 \times \pi \times 50$ = fundamental angular frequency

All outputs from the PMU are accurate at least up to 13 digits after decimal point for any input with nominal frequency and RMS magnitude up to 11kV or 11kA, as shown in Table I. Inputs greater than 11kV or 11 kA can be first scaled down to lower value before giving to PMU to preserve same order of accuracy.

TABLE I. PMU OUTPUTS FOR INPUTS WITH NOMINAL FREQUENCY

| PMU Outputs | | |
|---|---|---|
| Quantity | Value | Unit |
| RMS Magnitude | 230 | V |
| Frequency | 50 | Hz |
| Phase Angle -1 | -30 | Degree |
| Phase Angle -2 | 0 | Degree |
| Phase Angle -3 | +30 | Degree |
| Phase Angle -4 | -180 | Degree |

B. *PMU outputs for off-nominal frequency inputs:*

The developed PMU model has been tested for 4 different off-nominal frequency inputs with same phase angle of 90º. These testing inputs are unknown to the PMU when they are fed to it. The PMU accurately determines the magnitude, phase angle and frequency of each unknown input phasor. The 4 inputs with phase angle of 90º and different off-nominal frequencies are:

Input 1: $V = 230 * \sqrt{2} \cos(2\pi \times 49.5t + \frac{\pi}{2})$

Input 2: $V = 230 * \sqrt{2} \cos(2\pi \times 49.7t + \frac{\pi}{2})$

Input 3: $V = 230 * \sqrt{2} \cos(2\pi \times 50.3t + \frac{\pi}{2})$

Input 4: $V = 230 * \sqrt{2} \cos(2\pi \times 50.7t + \frac{\pi}{2})$

The frequency and RMS magnitude outputs from the PMU are accurate at least up to 14 digits after decimal point for any input with any off-nominal frequency and any RMS magnitude. The phase angle output by PMU, as per the property of recursive DFT technique, rotates at the rate determined by difference between input frequency and nominal frequency (50 Hz). The time period for a complete rotation of the phase angle from -180 to 180 is determined by (18), in which, if $f_{in}<f_0$ the phasor rotates clock-wise and if $f_{in}>f_0$ phasor rotates anti-clock-wise. Using (18), the periods for a complete rotation of the phase angles from -180 to 180 are calculated in Table II.

TABLE II. PERIODS OF PHASOR ROTATION FOR DIFFERENT OFF-NOMINAL FREQUENCY INPUTS

| Period of Phasor Rotation | | |
|---|---|---|
| Frequency (Hz) | T (s) | Direction |
| 49.5 | 2 | clock-wise |
| 49.7 | 3.3333 | clock-wise |
| 50.3 | 3.3333 | anti-clock-wise |
| 50.5 | 2 | anti-clock-wise |

The Fig. 1 shows phase angle outputs from PMU for the 4 off-nominal frequency inputs given.

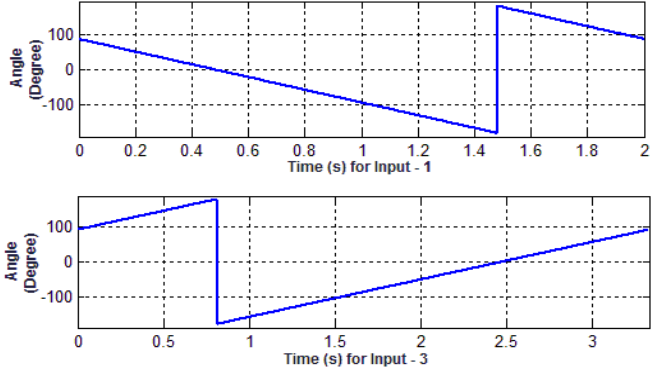

Fig. 1. Phase Angle Outputs for Off-Nominal Frequency Inputs

The following Table III shows magnitude and frequency outputs from the PMU for given 4 inputs as above.

TABLE III. RMS MAGNITUDE AND FREQUENCY OUTPUTS FOR OFF-NOMINAL FREQUENCY INPUTS

| PMU Outputs | | |
|---|---|---|
| Quantity | Value | Unit |
| RMS Magnitude | 230 | V |
| Frequency -1 | 49.5 | Hz |
| Frequency -2 | 49.7 | Hz |
| Frequency -3 | 50.3 | Hz |
| Frequency -4 | 50.7 | Hz |

## V. STEADY STATE SYNCHROPHASOR MEASUREMENT REQUIREMENTS [13, 14]

The compliance of the developed PMU model is checked against IEEE standard C37.118.1 [13, 14]. The standard proposes total 6 tests with pre-defined reference conditions. For any test, the error between estimated phasor and true phasor is determined by the criterion of total vector error (TVE) [13], which can be given as:

$$TVE(n) = \sqrt{\frac{(\hat{X}_r(n) - X_r(n))^2 + (\hat{X}_i(n) - X_i(n))^2}{(\hat{X}_r(n))^2 + (\hat{X}_i(n))^2}} \quad (26)$$

Where $\hat{X}_r(n)$ & $\hat{X}_i(n)$ denote the real and imaginary parts of the phasor estimate given by PMU, and $X_r(n)$ & $X_i(n)$ denote the real and imaginary parts of the true phasor at any time instant (n).

During the testing of the PMU model for steady state compliance, all influential quantities are kept constant for the period of measurement, including $X_m$, $\varphi$ & $\omega$ of the test signal. For the test signals with off-nominal frequencies, while input phase $\varphi$ is constant, the measured phase angle will keep on rotating as described by (18). The requirements for steady state compliance testing, as described in the standard [13], are given in Table IV.

TABLE IV. STEADY-STATE SYNCHROPHASOR MEASUREMENT REQUIREMENTS [13]

| Influence Quantity | Reference Condition | Minimum Range of Influence Quantity over which PMU shell be within given TVE Limit | | | |
|---|---|---|---|---|---|
| | | P Class | | M Class | |
| | | Range | Max TVE (%) | Range | Max TVE (%) |
| Signal Frequency Range: $f_0 \pm$ f_deviation | F_nominal ($f_0$) | ± 2.0 Hz | 1 | ±2.0Hz for $F_s$<10, ±$F_s$/5 for 10≤ $F_s$ <25, ± 5.0 Hz for $F_s$ ≥25 | 1 |
| Voltage Signal Magnitude | 100% (Rated) | 80% to 120% of Rated | 1 | 10 to 120% of Rated | 1 |
| Current Signal Magnitude | 100% (Rated) | 10% to 200% of Rated | 1 | 10 to 200% of Rated | 1 |
| Phase Angle Variation with $|f_{in}-f_0|$<0.25 Hz (see NOTE 1) | Constant or Gradually Varying Angle | ± π radians | 1 | ± π radians | 1 |
| Harmonic Distortion (Single Harmonic) | THD<0.2% | 1% of Each Harmonic up to 50th | 1 | 10%, Each Harmonic up to 50th | 1 |
| Out of Band Interference as described below (see Notes 2 & 3) | <0.2% of Input Signal Magnitude | <0.2% of Input Signal Magnitude | None | No requirement for $F_s$< 10. <10% of Input Signal Magnitude for $F_s$≥10. | 1.3 |

**Notes:**
**Out of band interference testing:** At each reporting rate, pass-band is termed as $|f - f_0|$< $F_s$/2. An interfering signal out-side pass-band of the filter is a signal with frequency f such that $|f - f_0|$≥ $F_s$/2. For this testing, $f_{in}$ - frequency of input test signal is changed between $f_0$ and ±10% of the Nyquist frequency of the reporting rate. i.e.:

$$f_0 - 0.1\left(\frac{F_s}{2}\right) \leq f_{in} \leq f_0 + 0.1\left(\frac{F_s}{2}\right) \quad (27)$$

Where $F_s$ = reporting rate of phasor,
$f_0$ = nominal frequency of the system,
$f_{in}$ = input test signal fundamental frequency.

**NOTE 1** — Phase angle test can be performed with $f_{in}$ - the input frequency off-set from $f_0$ where $|f_{in} - f_0|$< 0.25 Hz, which provides a gradually changing phase angle simplifying verification of compliance without producing other significant effects.

**NOTE 2** — A signal with frequency exceeding Nyquist rate for $F_s$ - the reporting rate can alias in the pass-band. The effectiveness of the PMU anti-alias filtering is verified by out of band interference testing signal which shall include frequencies causing the greatest TVE, outside of the bandwidth specified above.

**NOTE 3** — By using a single frequency sinusoid added to the fundamental power signal at a particular amplitude level, compliance with out of band rejection can be verified. Frequency of the signal is changed over a wide range from below the pass-band (as low as 10 Hz) and from above the pass-band to as high as 2<sup>nd</sup> harmonic level (2×f<sub>0</sub>). Interfering signal is a positive sequence, if the positive sequence measurement is being tested.

**NOTE 4** — The 2 performance classes – P class and M class are intended for fast response and optimum measurement respectively. Their detailed definitions can be found in the IEEE standard C37.118.1 [13, 14]. The PMU model developed here is generalized for either kind of application.

## VI. RESULTS OF STEADY-STATE COMPLIANCE TESTING

The tests described in IEEE standard are carried out for the developed PMU model with specified testing and reference conditions. The results obtained are given below in tabular forms.

### A. Frequency Test Results: 230 V RMS, constant angle -2π/3

For both P class and M class frequency tests, the maximum TVE is less than 1% except a particular frequency of 53 Hz, where max. TVE= 1.8%, as shown in Table V and Table VI.

TABLE V. P CLASS FREQUENCY TEST RESULTS

| Input Frequency | Maximum TVE |
|---|---|
| 48 | 0.0004 |
| 49 | 0.0004 |
| 50 | 0.0000 |
| 51 | 0.0004 |
| 52 | 0.0004 |

TABLE VI. M CLASS FREQUENCY TEST RESULTS

| Input Frequency | Maximum TVE |
|---|---|
| 45 | 0.0010 |
| 46 | 0.0017 |
| 47 | 0.0098 |
| 48 | 0.0004 |
| 49 | 0.0004 |
| 50 | 0.0000 |
| 51 | 0.0004 |
| 52 | 0.0004 |
| 53 | 0.0180 |
| 54 | 0.0017 |
| 55 | 0.0010 |

### B. Magnitude Test Results: at 230 V RMS, angle -pi/3

Since the PMU model operates identically in digital form to both current and voltage, tests 2 & 3 in standard are carried out as a single test for the maximum testing range of 10% to 200% of the nominal voltage value. The maximum TVE is 0.0000% for all tests, as shown in Table VII.

TABLE VII. P CLASS & M CLASS MAGNITUDE TEST RESULTS

| Magnitude as Percentage of Nominal Value | Maximum TVE |
|---|---|
| 10 % | 1.0e-13 *0.3792 = 0.0000 |
| 30 % | 1.0e-13 *0.3656 = 0.0000 |
| 50 % | 1.0e-13 *0.3608 = 0.0000 |
| 70 % | 1.0e-13 *0.3707 = 0.0000 |
| 90 % | 1.0e-13 *0.3611 = 0.0000 |
| 110 % | 1.0e-13 *0.3505 = 0.0000 |
| 130 % | 1.0e-13 *0.3517 = 0.0000 |
| 150 % | 1.0e-13 *0.3517 = 0.0000 |
| 170 % | 1.0e-13 *0.3601 = 0.0000 |
| 190 % | 1.0e-13 *0.3700 = 0.0000 |
| 200% | 1.0e-13 *0.3675 = 0.0000 |

### C. Phase Angle Test Results at 230 V RMS

Nominal Angle= – 90 degree= $-\frac{\pi}{2}$ radian. The maximum TVE is 0.0000% for all phase angles, as shown in Table VIII.

TABLE VIII. PHASE ANGLE TEST RESULTS

| Phase Angle Variations added to Nominal Value | Maximum TVE |
|---|---|
| $-\pi$ | 1.0e-13 *0.3586 = 0.0000 |
| $-\pi + \frac{\pi}{4}$ | 1.0e-13 *0.3954 = 0.0000 |
| $-\pi + 2\frac{\pi}{4}$ | 1.0e-13 *0.7056 = 0.0000 |
| $-\pi + 3\frac{\pi}{4}$ | 1.0e-13 *0.4029 = 0.0000 |
| $-\pi + 4\frac{\pi}{4}$ | 1.0e-13 *0.6577 = 0.0000 |
| $-\pi + 5\frac{\pi}{4}$ | 1.0e-13 *0.6623 = 0.0000 |
| $-\pi + 6\frac{\pi}{4}$ | 1.0e-13 *0.4325 = 0.0000 |
| $-\pi + 7\frac{\pi}{4}$ | 1.0e-13 *0.6315 = 0.0000 |
| $-\pi + 8\frac{\pi}{4} = \pi$ | 1.0e-13 *0.5718 = 0.0000 |

### D. Out of Band Interference Test: 230 V RMS, angle -π/4:

For the following two tests, the maximum TVE is less than 1%.
*1) Reporting rate 10 frames per second, input frequency within [49.5, 50.5] Hz, results are shown in Table IX.*

TABLE IX. 10 FPS OUT OF BAND INTERFERENCE TEST RESULTS

| Out of Band Interference Frequency | Maximum TVE |
|---|---|
| 49.5000 | 0.0000 |
| 49.6250 | 0.0052 |
| 49.7500 | 0.0000 |
| 49.8750 | 0.0027 |
| 50.0000 | 0.0000 |
| 50.1250 | 0.0030 |
| 50.2500 | 0.0002 |
| 50.3750 | 0.0038 |
| 50.5000 | 0.0003 |

*2) Reporting rate 25 frames per second, input frequency within [48.75, 51.25] Hz, results are shown in Table X.*

TABLE X. 25 FPS OUT OF BAND INTERFERENCE TEST RESULTS

| Out of Band Interference Frequency | Maximum TVE |
|---|---|
| 48.7500 | 0.0000 |
| 49.0000 | 0.0000 |
| 49.2500 | 0.0005 |
| 49.5000 | 0.0000 |
| 49.7500 | 0.0000 |
| 50.0000 | 0.0000 |
| 50.2500 | 0.0002 |
| 50.5000 | 0.0003 |
| 50.7500 | 0.0012 |
| 51.0000 | 0.0006 |
| 51.2500 | 0.0008 |

## VII. ANALYSIS OF THE RESULTS

### A. Limitations of the developed PMU Model:

1) In frequency test, the PMU gives max. TVE=1.8% for a particular frequency of 53 Hz, while max. TVE<1% for all other frequencies. This can be taken care of by modifying the algorithm for the particular frequency of 53 Hz.
2) The PMU model has been developed for optimum performance at fundamental frequency ($f_0$). It fails to operate for frequencies $\geq 2*f_0$ and so it is unable to detect harmonic content present in input signal. This is due to the theoretical constraint of the equations employed in recursive DFT algorithm which limits the input frequency $f_{in} < 2*f_0$ to avoid time-domain aliasing effects due to harmonics.

### B. Solution to the limitations of the PMU Model:

To detect harmonic content other than fundamental, additional frequency spectrum analyzer can be incorporated as a hardware unit. The input should first pass through spectrum analyzer and then each harmonic component should be given to independent PMU intended to consider corresponding harmonic frequency as the fundamental one, ensuring same accuracy for each harmonic measurement.

## VIII. CONCLUSION

The developed PMU model has been tested as per IEEE standard C37.118.1 and it is concluded that the model gives quite accurate results for steady state operation and it can be recommended for practical implementation. This steady state model can be incorporated with existing models optimized for dynamic conditions, in order to get most accurate results in both steady state and dynamic states.

ACKNOWLEDGEMENT

The author would like to thank Marco Pau, Ph.D., E.ON Energy Research Centre, RWTH Aachen University, Germany for his suggestion and guidance for research in this field. The author is grateful to Dr. Prasanta Kundu, Assistant Professor, EED, SVNIT for providing necessary infrastructure and guidance.